# Band Structure and Polarization Effects in Photothermoelectric Spectroscopy of a Bi$_2$Se$_3$ Device

Seyyedesadaf Pournia,[1] Giriraj Jnawali,[1] Ryan F. Need,[2,3] Howard E. Jackson,[1] Stephen D. Wilson,[2] and Leigh M. Smith,[*,1]

[1] *Department of Physics and Astronomy, University of Cincinnati, Cincinnati, OH 45221, USA.*
[2] *Materials Department, University of California, Santa Barbara, CA 93106, USA*
[3] *Department of Materials Science and Engineering, University of Florida, Gainesville, FL 32611, USA*

*Author to whom correspondence should be addressed: leigh.smith@uc.edu

## Abstract

Bi$_2$Se$_3$ is a prototypical topological insulator which has a small band gap (~0.3 eV) and topologically protected conducting surface states. This material exhibits quite strong thermoelectric effects. Here we show in a mechanically exfoliated thick (~100 nm) nanoflake device that we can measure the energy dependent optical absorption through the photothermoelectric effect. Spectral signatures are seen for a number of optical transitions between the valence and conduction bands, including a broad peak at 1.5 eV which is likely dominated by bulk band-to-band optical transitions but is at the same energy as the well known optical transition between the two topologically protected conducting surface states. We also observe a surprising linear polarization dependence in the response of the device that reflects the influence of the metal contacts.

Finding ways to obtain sensitive measures of the absorption as a function of wavelength in nanostructured materials is a key to their characterization. While absorption is a standard technique for bulk single crystals, it is challenging to make such measurements in nanostructures. Reflectivity is one possibility but it requires tuning over very large wavelength ranges in order to accurately convert to absorption through a Kramers-Kronig relation. Including the effects of the substrate is particularly difficult. For strongly emitting nanostructures, photoluminescence excitation (PLE) is possible, but this is only an approximation since it depends on whether or how strongly the excited states are coupled to the emitting state. Here we demonstrate in the trichalcogenide compound $Bi_2Se_3$ that one can obtain a direct and sensitive measure of the absorption using photothermoelectric spectroscopy.

The trichalcogenide compound $Bi_2Se_3$ has long been known as a thermoelectric material with the ability to convert heat into electricity.[1–6] $Bi_2Se_3$ is also known as a prototypical topological insulator with nontrivial surface states with unique optical effects such as the circular photogalvanic effect (CPGE) and the photoinduced inverse spin Hall effect.[7–18] Underlying all of that, bulk $Bi_2Se_3$ is a standard semiconductor that possesses allowed optical transitions between the occupied valence bands to the unoccupied conduction bands.[19]

In this letter we show that the photoresponse of a single $Bi_2Se_3$ nanoflake device is dominated by the photothermoelectric effect. The photothermoelectric effect occurs when a light beam is absorbed at one spatial position in a device, causes a temperature gradient across the sample, and finally generates a voltage through the Seebeck effect. The resulting signal is directly proportional to the absorbed power, which is directly proportional to the absorption itself. We use a laser which can be tuned from 0.3 to 1.8 eV to measure directly the absorption of a single $Bi_2Se_3$ nanoflake as a function of energy and observe a series of optical transitions between the valence and conduction bands. We demonstrate that the metallic contacts of the device cause measurable polarization effects which are predicted through FDTD calculations. The results shown here demonstrate the ability to measure the energy dependent absorption of any nanostructure which exhibits thermoelectric effects, even gapless semimetals.

Single crystals of $Bi_2Se_3$ were grown using the Bridgeman method from stoichiometric elements as described elsewhere.[19] The bulk crystals were degenerately doped n-type ($N_D \sim 10^{19}$ cm$^{-3}$). Relatively thick (~100 nm) nanoflakes were exfoliated using the scotch tape method and transferred to a high purity p-type silicon substrate with a 300 nm thermal oxide layer. After visualization of the nanoflakes using a microscope, metal contacts with ~5 $\mu m$ gap were defined across the nanoflakes using photolithography followed by e-beam deposition of a Ti/Al thin film and liftoff. After confirming that the contacts were ohmic using I-V measurements in a probe station, the device was mounted into a chip carrier and wires bonded to the contacts. The assembly was mounted to the copper cold finger in the vacuum space of a ST-500 Janis microscope cryostat with an Attocube piezoelectric xy-scanner for fine positioning and a thin $CaF_2$ infrared window for conducting the optical experiments. The temperature could be varied between 10 K and 300 K. Transport measurements were implemented using a SRS variable constant current source and measuring the change in the bias.

The tunable output of an optical parametric oscillator pumped by a pulsed (150 fs/80 MHz) Ti-sapphire laser was used to photoexcite the device over a continuous energy range extending from 0.3 eV (4020 nm) to 1.8 eV (680 nm). A 40×/0.5 N.A. protected silver reflective objective was used to focus the laser beam to a diffraction-limited spot onto the sample (~1 to 5 μm, depending on wavelength). The average power could be varied between 150 and 750 μW ($P_0/5$ and $P_0$, respectively), and the orientation of the linear polarization relative to the contacts could be controlled using a $CaF_2$ double Fresnel rhomb rotator.

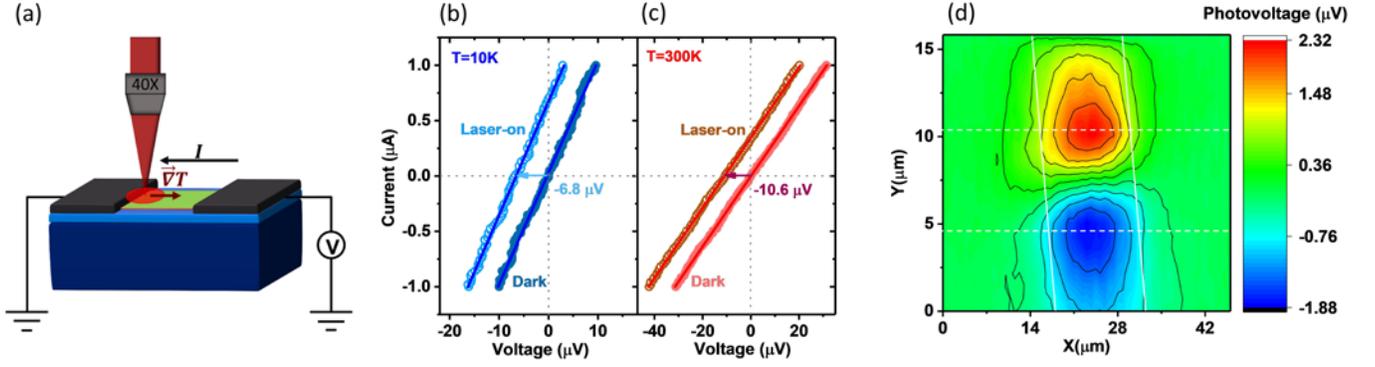

**FIG. 1.** (a) Schematic diagram of the experiment. I-V measurements at (b) 10K and at (c) 300K in the dark and under 1.16 eV (1070 nm) illumination with power of $P_0$. (d) 2D mapping of the device photoresponse with a power of $P_0/5$ at 300 K. Solid lines show approximate edges of Bi$_2$Se$_3$ nanoflake, while dashed lines show approximate edges of metallic contacts.

Figure 1(a) shows a schematic diagram of a photovoltage measurement in a Bi$_2$Se$_3$ nanoflake device. With the laser positioned close to one contact, current-voltage (I-V) measurements are shown at 10 K (Fig. 1(b)) and 300 K (Fig. 1(c)) in the dark and under 1.16 eV (1070 nm) laser illumination. The I-V measurements are clearly linear indicating that the metal-semiconductor contacts are ohmic. The resistance of the device is 31 Ω at 300 K and 10 Ω at 10 K which, since the resistance of the device decreases with temperature, exhibits metallic behavior. Given the thickness of the flake (~ 100 nm) with the distance between the contacts (~5 μm) and the width of the flake (~15 μm), we calculate that the resistivity of the Bi$_2$Se$_3$ is $1.6 \times 10^{-3}$ Ω.cm at 300 K and $5 \times 10^{-4}$ Ω.cm at 10 K. Considering the average literature values for mobility in similar n-type (metallic) Bi$_2$Se$_3$ devices, $\mu_{300K} = 550$ cm$^2$/V.s and $\mu_{10K} = 1800$ cm$^2$/V.s,[20–22] we estimate that the free carrier density to be $n_{3D} = 7 \times 10^{18}\ cm^{-3}$, which is in agreement with the result of analysis of the transient reflectivity (TR) spectra taken on a nanoflake exfoliated from the same bulk crystal.[19] This carrier density indicates a highly doped sample, which has been seen in similar crystals grown using the growth method described previously.[19]

Under illumination with 750 µW of 1.16 eV (1070 nm) light, the slope of the I-V curve does not change demonstrating that the resistivity does not change under illumination. If we assume all of the light is absorbed by the flake, the expected photoexcited pair density is less than 10% of the background electron density in steady state assuming an average lifetime of 150 ps and a spot size of ~2 µm.[19] On the other hand, the I-V profile is shifted by 6.8 µV at 10 K and by 10.6 µV at 300 K from the measurements in the dark. This rigid shift indicates that while the conductivity does not change, a voltage across the contacts is induced by the illumination of the laser. Assuming that the power is completely absorbed near one contact (*cf.* Fig. 1(a)) the laser creates a temperature gradient across the device which in turn induces a voltage across the contacts through the Seebeck effect. This phenomena is known as the photothermoelectric effect (PTE),[23,24] in which illumination drives current flow under short circuit (V = 0) conditions, or creates a potential difference between the two electrodes ($\Delta V_{PTE}$) under open circuit (I = 0)[23] conditions. The magnitude of the photothermoelectric effect depends on the Seebeck coefficient, which in turn depends on the conductivity of the sample near the Fermi energy, and the temperature gradient induced between the two ends of the material. The temperature gradient in steady state is determined by the amount of energy absorbed at one end, and the thermal conductivity of the material.

The conclusion that PTE drives the photoresponse is supported by the photovoltage map shown in Figure 1(d) which displays how the induced voltage at constant current (I = 0) varies as a function of the position of the laser. The dotted lines show the edges of the metal contacts and solid lines show the edges of the $Bi_2Se_3$ nanoflake. The power of the 1.16 eV laser was $P_0/5 = 150$ µW with a spot size of 1.4 µm in this image determined by the numerical aperture (0.5) of the reflective objective. The maximum induced voltage is seen when the laser is positioned at the edge of the metal contacts which would be expected to create the largest temperature gradient across the flake. The voltages are positive when the laser is near one contact and negative when it is near the other because the temperature gradient reverses sign. Prior measurements of similar $Bi_2Se_3$ material have shown the Seebeck coefficient to range from $51 - 82$ μV/K,[25–27] which suggests a ~ 34 mK temperature difference between the two ends of the flake, for $\Delta V_{PTE} = 2.32$ μV (Fig. 1(d)) and assuming all of the incident $P_0/5 = 150$ μW is absorbed by the material. The induced voltage midway between the contacts is zero since there is no temperature gradient in that case.

Interestingly, the induced voltage is relatively constant as a function of position when the flake is illuminated through the aluminum contacts. This indicates that the absorbed power does not change appreciably when the laser is over the contacts, but also indicates that the temperature is relatively constant for the flake underneath the contacts (with a negligibly small temperature gradient), resulting in a temperature gradient which is maximum between the contacts. The small thermal gradient under the contacts is consistent with the high thermal conductivity of the aluminum relative to the underlying $Bi_2Se_3$ resulting in a nearly constant temperature no matter where the laser spot is located on the metal film.

At constant current (here I = 0), the potential induced by PTE depends on the induced temperature gradient between the contacts, which depends on the amount of laser power absorbed by the nanoflake. It should thus be possible to measure how the absorption changes in the $Bi_2Se_3$ with energy. Figure 2(b) shows the induced photovoltage in the device as a function of photon energy from 0.3 to 1.8 eV (4020 nm – 680 nm) at 300 K (red spectrum) and 10 K (blue spectrum). The laser was focused on a point close to the metal contact with the maximum photoresponse. Here, the photoresponse of the device has been normalized by the power of the laser, $(\Delta V_{PTE}/P)$, for each single wavelength in order to obtain a measure of how the absorption changes with wavelength. The photovoltage spectra in Fig. 2(b), shows three peaks at 0.36 eV, 1.1 eV and 1.51 eV noted by blue bands. A schematic of the band structure of the material and carrier transitions related to the energy peaks of the spectra is depicted in Fig. 2(a). After a rapid onset beginning at 0.32 eV, the peak at 0.36 eV is the optical band gap determined by optical transitions between the uppermost valence band ($VB_1$) and the Fermi level ($E_f$) in the conduction band (pink arrow) determined by the electron doping density. Previous transient reflectivity measurements in a flake exfoliated from the same single crystal have shown from spectral analysis that the fundamental gap is ~0.2 eV with a Fermi energy of 120 meV which matches the onset observed here at 0.32 eV.[19] The second peak at 1.1 eV, is the energy required to excite the carriers from the second valence band ($VB_2$) to the Fermi level (green arrow) and was also resolved in the TR measurements on a similar nanoflake.[19] The third broad peak centered at 1.51 eV (820 nm) is likely dominated by bulk band-to-band optical transitions, but is also at the same energy as the well-known optical transition between the first to second Dirac surface states (Fig. 2(a), blue arrows).[28] This SS1 to SS2 optical transition has been confirmed experimentally[29] in surface sensitive measurements and most CPGE measurements on this TI material are performed at this energy.[13,16]

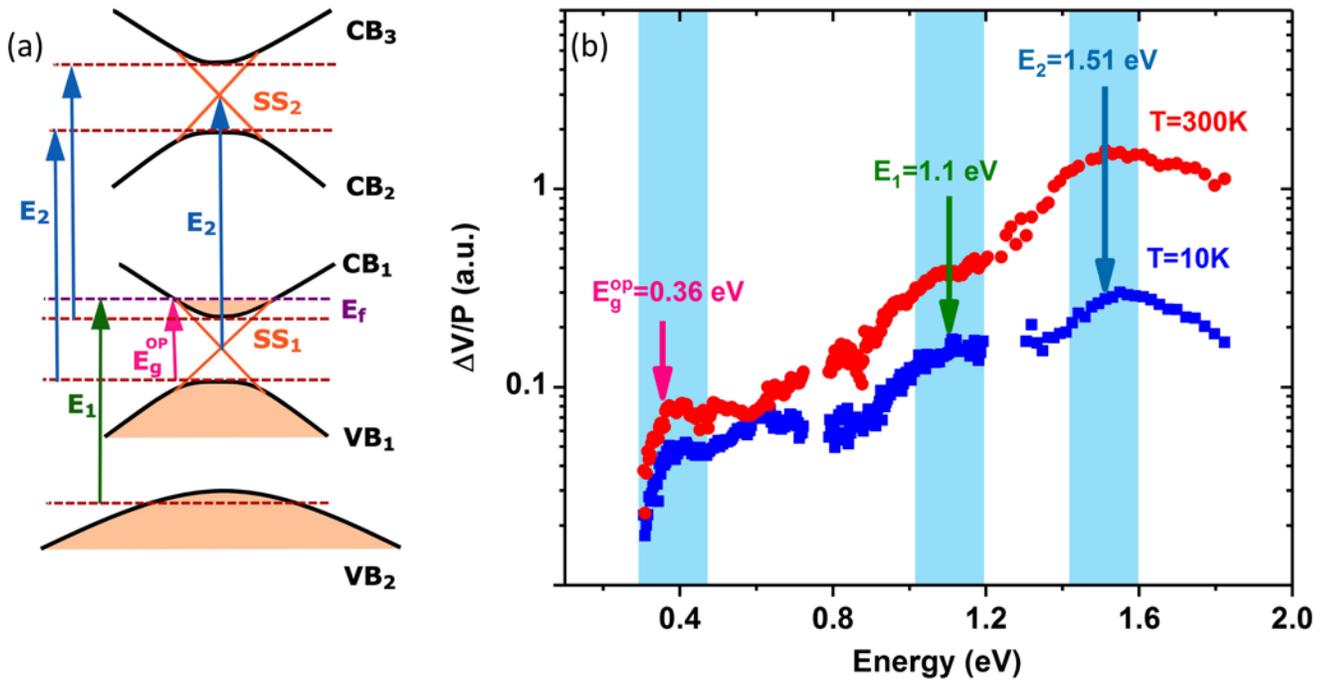

**FIG. 2.** (a) Schematic of the Bi$_2$Se$_3$ energy band structure with optical transitions noted by vertical arrows. (b) The induced photovoltage spectra at 300K (red) and 10K (blue). Spectral features are shown by shaded vertical bars noting the optical transitions for (E$_g$) the optical gap, (E$_1$) the transition from the second VB to the Fermi level in the CB, and (E$_2$) the transitions between different bulk states and from the first Dirac surface state SS1 to second Dirac surface states SS2.

The energy-dependent measurements in Fig. 2 are consistent with the expected dependence of the absorption in this material. Three potential sources for the measured response include the photothermoelectric effect, the photobolometric effect and the photovoltaic effect.[30] The photobolometric effect occurs because the transport coefficients change as a function of temperature. In Fig. 1(b) the temperature dependent data implies that the temperature sensitivity of the resistance from the IV measurements reduces by a factor of three from 300 K to 10 K or ~1%/K. We observe that the slopes of the IV curves measured in Fig. 1(b) change by less than 1% when illuminated by 150 µW of 1070 nm light suggesting that the temperature change is less than 1 K at these small illumination powers. We conclude the photobolometric effect would be only 0.02 µV/K, orders of magnitude smaller than the Seebeck coefficient which is ~68 µV/K. Photovoltaic signals occur because of the presence of Schottky barriers from the metal contacts would cause electric fields which separate electrons and holes. As shown in Fig. 1 (b) and (c) the IV curves are extremely linear suggesting that the contacts are ohmic. More importantly, the spatial map shown in Fig. 1 (d) was taken at 1070 nm illumination with an approximate spatial resolution of one micron, and yet the photovoltage signal taken under open circuit conditions extends well beyond the edge of the metal contact. This is inconsistent with a photovoltaic response which would exhibit a dominant response at the edge of the contact limited by the spatial spot size of the laser. We conclude that the measured signal is dominated by the photothermoelectric effect in this device.

We now discuss how we can understand this detection mechanism based on the Seebeck effect. The temperature gradient induced by the laser ($dT/dx$) can be determined by Eq. 1 which shows that the rate at which heat is deposited in the sample ($dQ/dt$) is balanced by the transport of heat away from the excitation point, which in turn is determined by the thermal conductivity, $\kappa$, and the cross-sectional area of the flake.[31] The absorbed power ($dQ/dt$) is also equal to the energy dependent absorption coefficient $\alpha(\mathcal{E})$ times the incident laser power $P$, leading to Eq. 2.

$$\frac{dQ}{dt} = -A\,\kappa\,\frac{dT}{dx} \qquad (1)$$

$$\alpha(E)P = -A\,\kappa\,\frac{dT}{dx} \qquad (2)$$

However, the PTE voltage is directly proportional to the temperature gradient through the Seebeck coefficient, $\Delta V_{PTE} = -S\,\Delta T$ and so $\Delta V/P \sim \alpha$. Thus, we can conclude that $\Delta V/P$ as a function of photon energy is a direct measure of the energy-dependent absorption. Simple estimates show that less than 1% of the absorbed power is actually transmitted through the $Bi_2Se_3$ nanoflake. Most of the thermal energy goes into the large aluminum contacts (the gold wires bonded to the pads define the background temperature), and through the underlying $SiO_2$/Si substrate. The PTE signal could thus be dramatically improved with simple changes to the device geometry (e.g., Using much smaller contacts to the nanoflake, separating further apart, or suspending the nanoflake over a trench).

Notably, we have found that the maximum photovoltage depends on the orientation of the laser polarization relative to the contact. This is unexpected since the absorption coefficient along the $a$ or $b$ crystallographic directions for $Bi_2Se_3$ should not depend on polarization (i.e., the response should be isotropic). For $Bi_2Se_3$ with rhombohedral crystal structure and a hexagonal surface, the absorption coefficients along $a$ and $b$ crystal axes should be the same by symmetry. Figure 3(a) shows the photovoltage under constant 1070 nm wavelength illumination as a function of laser polarization. The induced photovoltage is clearly maximum for a polarization perpendicular to the electrode (parallel to the current), which means that the temperature gradient is larger which means the absorbed power is also larger. We can fit the polarization dependence of the photovoltage as $\Delta V = D + L\cos(2\varphi)$, where $D$ is the polarization independent response, $L$ is the polarization dependent term which is equal to about 15 % of $D$, and $\varphi$ is the polarization angle with respect to perpendicular to the contact. The red (blue) profile is for the measurements close to the top (bottom) contact in Fig. 1(d). This result clearly shows that the positive or negative current is optimized for the polarization aligned with the current, and is about 25% less for the polarization perpendicular to the current or parallel to the contacts.

$Bi_2Se_3$ exhibits a polarization dependent linear photogalvanic effect (LPGE) but this occurs only for laser illumination not at normal incidence but at an angle. Furthermore, it should not depend on the spatial position of the laser. We do not see any polarization dependence (or signal) midway between the contacts where the PTE signal is zero. We conclude that the polarization dependence must come from the contacts.

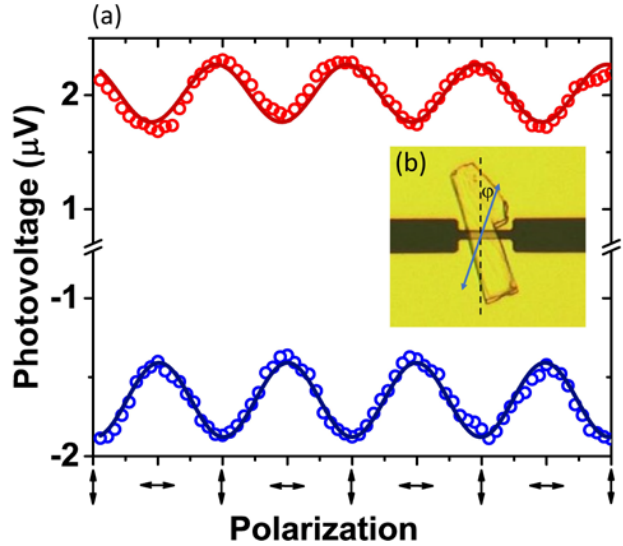

**FIG. 3.** (a) Linear polarization dependence of the photothermoelectric (PTE) response. (b) Optical image of the $Bi_2Se_3$ device used in this experiment and the schematic polarization vector orientation with respect to the normal of the contact edges.

To support this conclusion, we simulated the anisotropic effect of the contacts using the FDTD calculator program Lumerical. As shown in Fig. 4, when 1070 nm laser illumination is focused well away from the contact (x = 1.5 $\mu m$ and y = 1.5 $\mu m$), the square of the electric field ($E^2$) has no polarization dependence. However, when the laser is focused directly on the edge of the contact (x = 1.5 µm and y = 4.5 µm), $E^2$ (and so $\alpha$) is larger for the linear polarization perpendicular to the metal contact (parallel to the induced current). This response can be completely understood by considering the boundary conditions for the electric field at the edge of a metal contact. For a perfect metal, the tangential electric field must be zero at the interface for continuity, while the perpendicular component of the electric field induces a surface charge which allows a discontinuous transition to zero field inside the metal. The transition is limited to the skin depth. The depletion of the electric field at the interface for parallel polarization is caused by the fact that it must go to zero inside the metal. Indeed, this depletion is seen as the electric field is measured within 200 nm of the interface, even though the laser spot (the maximum of the incident intensity) is located directly on the interface. On the other hand, for perpendicular polarization, the field strength is also clearly perturbed causing a slight 8% strengthening of the field intensity at the interface, and a more rapid falloff of the strength with distance. Both of these effects are responsible for the observed polarization dependence. One can estimate the degree of linear dichroism to be ~25% by estimating the maximum intensities for each polarization from the FDTD calculation. This is consistent with the measured value of 25%.

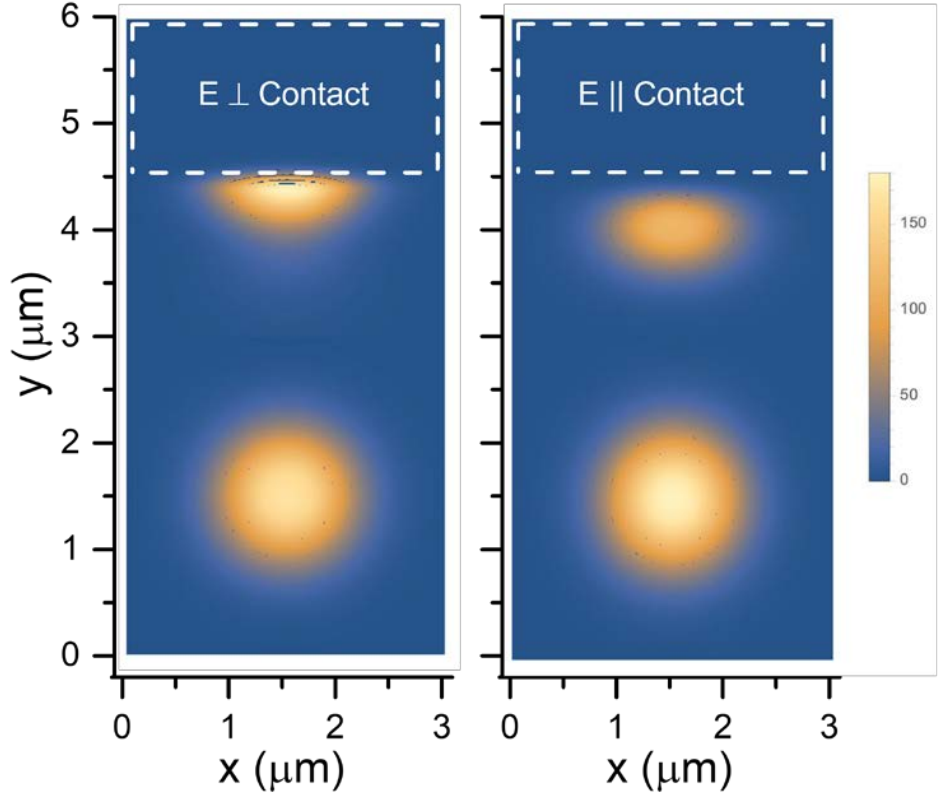

**FIG. 4.** FDTD Calculation of the square of the electric field $E^2$ for a linearly polarized 1070 nm laser beam focused to a spot on a $Bi_2Se_3$ nanoflake device. The left panel shows the field intensity for incident light polarized perpendicular to the Ti/Al contact, while the right panel shows the field intensity for light polarized parallel to the contact. When the spot is focused well away from the contact the field intensities for each polarization are the same. When light is focused on the Ti/Al metallic contact the intensity is enhanced for perpendicularly polarized light and depleted for parallel polarized light.

In summary, we have shown that it is possible to use the photothermoelectric effect to directly measure the energy dependence of the absorption in a $Bi_2Se_3$ nanoflake device. We see spectral features that are associated with well-known allowed optical transitions between the valence bands and the conduction bands and are consistent with transient reflectivity measurements on the same material.[19] In addition, we see a peak in the response at 1.5 eV which is likely a combination of bulk optical transitions and direct optical transitions between the two topologically protected surface states in this well-known topological insulator. While the symmetry of $Bi_2Se_3$ suggests it should not show any absorption polarization dependence in our geometry, we find a measurable linear dichroism for light which is focused on the metal contact of the device. We show through direct FDTD calculations that this linear dichroism effect is caused by the expected boundary conditions at the contact interface, which slightly enhances the absorption of the perpendicular polarization and strongly depletes the parallel polarization. Designs of future devices which involve electrodes need to be aware of these significant polarization effects. If devices can be designed where the photoresponse is dominated by the photothermoelectric effect, the results described here open the possibility of making extremely sensitive measurements of absorption in conductive semiconductor or semimetal nanostructures in the future.

We acknowledge the financial support of the NSF through grants DMR 1507844, DMR 1531373, and ECCS 1509706. SDW acknowledges support via the UC Santa Barbara NSF Quantum Foundry funded via the Q-AMASE-i program under award DMR-1906325.

**Data Availability**

The data that support the findings of this study are available from the corresponding author upon reasonable request.